\documentclass[twocolumn,prl]{revtex4-2}
\usepackage{natbib}
\usepackage{ulem}
\usepackage{graphicx, epsfig}
\usepackage{latexsym}
\usepackage{epsfig}
\usepackage{amstext}
\usepackage{ulem}
\usepackage{amssymb}
\usepackage{array}
\usepackage{color}
\usepackage{graphicx}
\usepackage{dcolumn}
\usepackage{amsmath}
\usepackage{amsfonts}
\usepackage{amsthm}
\usepackage{newlfont}
\usepackage{graphics}
\usepackage{mathrsfs}
\usepackage{hyperref}
\usepackage{eucal}
\usepackage{verbatim}
\usepackage{fancyhdr}
\usepackage{bm}
\usepackage{lineno}
\usepackage{xcolor}

\newcommand{\p}{\partial}

\newcommand{\eps}{\varepsilon}
\newcommand{\x}{\mbox{\boldmath$x$}}

\newcommand{\y}{\mbox{\boldmath$y$}}

\newcommand{\n}{\mbox{\boldmath$n$}}

\newcommand{\FF}{\mathcal{F}}
\newcommand{\TT}{\mathcal{T}}

\newcommand{\RR}{\mathcal{R}}

\definecolor{red}{rgb}{1,0,0}

\begin{document}
\title{Voltage mapping in subcellular nanodomains using electro-diffusion modeling}
\author{Frédéric Paquin-Lefebvre$^{1}$ and David Holcman$^{1,2}$}
\email[Corresponding author: ]{david.holcman@ens.fr}
\affiliation{$^{1}$Group of Data Modeling, Computational Biology and Applied Mathematics, \'Ecole Normale Sup\'erieure -- Université PSL, 75005 Paris, France. \\
$^{2}$Department of Applied Mathematics and Theoretical Physics and Churchill College, University of Cambridge, Cambridge CB3 0WA, United Kingdom.}
\begin{abstract}
\noindent Voltage distribution in sub-cellular micro-domains such as neuronal synapses, small protrusions or dendritic spines regulates the opening and closing of ionic channels, energy production and thus cellular homeostasis and excitability. Yet how voltage changes at such a small scale in vivo remains challenging due to the experimental diffraction limit, large signal fluctuations and the still limited resolution of fast voltage indicators. Here, we study the voltage distribution in nano-compartments using a computational approach based on the Poisson--Nernst--Planck equations for the electro-diffusion motion of ions, where inward and outward fluxes are generated between channels. We report a current--voltage (I--V) logarithmic relationship generalizing Nernst law that reveals how the local membrane curvature modulates the voltage. We further find that an influx current penetrating a cellular electrolyte can lead to perturbations from tens to hundreds of nanometers deep depending on the local channels organization. Finally, we show that the neck resistance of dendritic spines can be completely shunted by the transporters located on the head boundary, facilitating ionic flow. To conclude, we propose that voltage is regulated at a subcellular level by channels organization, membrane curvature and narrow passages.
\end{abstract}
\maketitle
\section{Introduction}
\begin{figure}[!ht]
\begin{center}
\includegraphics[width=\linewidth,height=10cm]{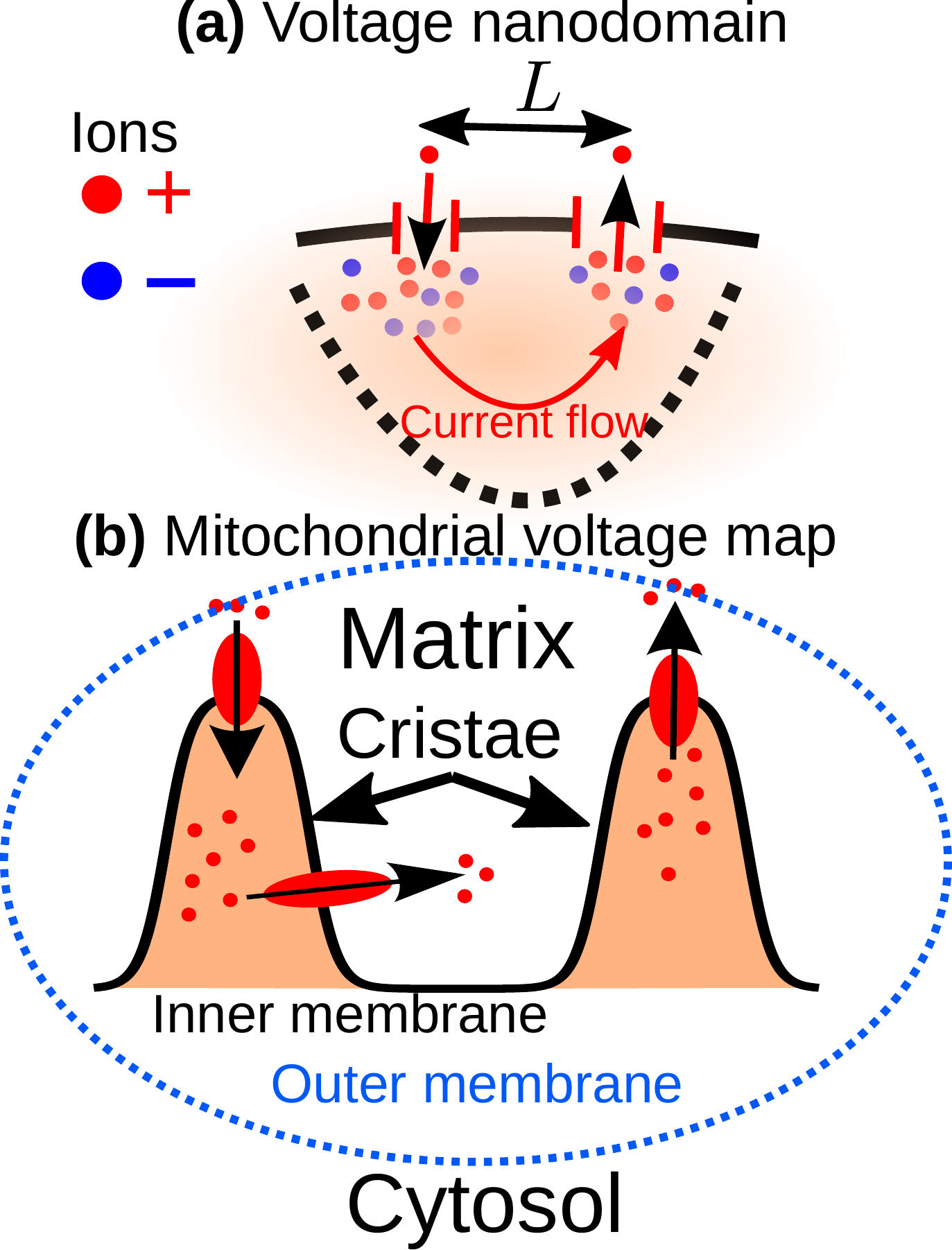}
\caption{\label{fig:fig1} \textbf{Ionic currents through channels generate voltage nanodomains.} (\textbf{a}) Generic model of a voltage nanodomain (blurred region) located in the vicinity of two ionic channels separated by a distance of $L$. In these models only positive charges can be exchanged. (\textbf{b}) The production of ATP in mitochondria requires the exchange of positive hydrogen ions across the inner membrane, a process that could be modulated by the curvature \cite{garcia,emboJ}.}
\end{center}
\end{figure}

Voltage changes in subcellular compartments regulate cell excitability and homeostasis \cite{hille,Turrigiano}, cell--cell communication and plastic changes allowing the storage of short- and long-term memory \cite{nicoll,Huganir,yuste2010dendritic}. For neuronal cells, the voltage is regulated by the opening and closing of a large variety of voltage-sensitive channels that can be specific to Na$^+$, K$^+$, Cl$^-$ and Ca$^{2+}$ ions, with pumps and exchangers allowing ionic flow outside of the cell. Upon selection, ions flow inside channels one by one until the cytoplasm \cite{bezanilla2008}, where they can spread in a three-dimensional structure below the channel pore, thereby affecting local ionic concentrations. For instance, following the opening of voltage-gated channels intracellular calcium concentration can increase from the nanomolar to the micromolar range, representing a factor of $\sim$1000 \cite{hille,zamponi2014}. In addition, ionic channels such as Ca$_v$--K$_v$ \cite{zamponi2011} are sometime activated together (co-activated): calcium channels open, which can trigger the opening of the nearby potassium channels located within tens to hundreds of nanometer distances. However, the exact distance between these channels that controls their opening-closing remains unknown, which motivated the present study. Note that the resulting local changes in concentrations lead to voltage changes that can modulate the opening probability of nearby channels through the formation of a voltage nanodomain. \\
Although voltage measurements in neuronal nanodomains remain difficult due to their small size, recent developments in voltage-sensitive dyes now allow a nanometer scale resolution \cite{Dieudonne} to measure neuronal spikes or map the voltage in mitochondria \cite{emboJ}. A complementary approach uses nanopipettes allowing a better temporal resolution while loosing the necessary spatial accuracy to measure the voltage in dendrites and dendritic spines \cite{Lagacheyuste,JeffHugh2023}, with the interpretation of these measurements still unclear. \\
Here, we investigate how the voltage spreads at a subcellular level following ionic current flow through nearby channels, leading to the formation of a voltage disturbed nanodomain. For this purpose we analyze an electro-diffusion model formulated with the Poisson--Nernst--Planck (PNP) equations {\cite{qian1989,mori2007,schussEisenberg2001,singer2008,pods2013}}. This approach allows us to compute the voltage directly from ionic densities, and is more general than the classical cable equation \cite{tuckwell1988,rallSegev} which ignores any spatial ionic density variations. Previous modeling and simulation reports revealed that the voltage distribution is influenced by concentration differences due to specific geometry such as the head-neck organization of dendritic spines \cite{rusakov2017,cartailler2018neuron}. In addition, while the Poisson--Boltzmann theory describes voltage distribution and local charge imbalances near planar membranes \cite{andelman,BenYaakov2009,orland2000}, it cannot be used to model ionic fluxes originating from channels and pumps. \\
Our computational framework is based on a two-charge model with a single positive ionic species flowing in and out of the domain, while negative charges remain inside [Fig.~\ref{fig:fig1}(\textbf{a})], and we present asymptotic and numerical results for the voltage organization between two channels. We obtain a logarithmic current--voltage (I--V) relationship that accounts for nonlinear effects at large current amplitudes and reveals the influence of the channels organization and the local membrane curvature. We then characterize how an influx current spatially extends within an electro-neutral medium by computing the deepest point along trajectories following the electro-diffusion gradient and report penetration lengths ranging from tens to hundreds of nanometers, in contrast with the Debye length \cite{andelman} of less than a few nanometers measuring the width of the non-neutral boundary layer near cellular membranes. We also show that voltage dynamics in cellular nanodomains can be modulated by the membrane curvature, which is the case for the inner mitochondrial membrane \cite{garcia,emboJ}, where the folding cristae structures imposes a large curvature [Fig.~\ref{fig:fig1}(\textbf{b})]. Finally, we analyze the voltage distribution in a dendritic spine micro-domain consisting of a ball connected to a narrow cylindrical neck. Following the injection of a synaptic current, the narrow neck acts as an electrical resistance \cite{cartailler2018neuron}, but interestingly, we show that by adding multiple exit sites to the head boundary the neck can be transformed into an effective conductor over its entire length.
\section{Voltage distribution computed from the electro-diffusion model} 
Ionic concentrations in an electrolyte within a domain $\Omega$ can be computed using the Poisson--Nernst--Planck equations \cite{hille,bazant}. For two monovalent ionic species of positive $c_p(\x,t)$ and negative $c_m(\x,t)$ charges, the motion results from the coupling between diffusion and advection due to the electric field, yielding the two fluxes
\begin{align}
j_p(\x,t) &= - D_p\hspace{-1mm}\left(\nabla c_p(\x,t) + \frac{ec_p(\x,t)}{k_B \TT}\nabla v(\x,t)\right), \\
j_m(\x,t) &= - D_m\hspace{-1mm}\left(\nabla c_m(\x,t) - \frac{ec_m(\x,t)}{k_B \TT}\nabla v(\x,t)\right),
\end{align}
where $D_p$ and $D_m$ are the diffusion coefficients, $k_B\TT/e$ is the thermal voltage and $v(\x,t)$ is the voltage variable. The Nernst--Planck equations are summarized by the conservation of positive and negative charges,
\begin{subequations}\label{eq:np}
\begin{align}
\frac{\p c_p(\x,t)}{\p t} + \nabla \cdot j_p(\x,t) &= 0\,, \quad \x \in \Omega\,, \\
\frac{\p c_m(\x,t)}{\p t} + \nabla \cdot j_m(\x,t) &= 0\,, \quad \x \in \Omega\,.
\end{align}
\end{subequations}
The voltage distribution is defined by ionic fluxes crossing the boundary $\p\Omega$ through two narrow circular windows $\p\Omega_{A_1}$ and $\p\Omega_{A_2}$, of radii $A_1$ and $A_2$, and centered at $\x_1$ and $\x_2$, respectively, with $L = \|\x_1-\x_2\|$ the distance in between (Fig.~\ref{fig:fig1}(\textbf{a})). The first window receives a current composed of positive charges while the second expels a flux, with the remaining boundary part $\p\Omega_r = \p\Omega \backslash \left( \p\Omega_{A_1} \cup \p\Omega_{A_2} \right)$ that is fully reflective of both ionic species. On window $\p\Omega_{A_1}$ a positive charge current of amplitude $I$ is injected, thus yielding
\begin{equation}\label{eq:influx}
\n \cdot j_p(\x,t) + \frac{I}{\FF \pi A_1^2} = 0, \,\, \n \cdot j_m(\x,t) = 0, \,\, \x \in \p\Omega_{A_1}\,,
\end{equation}
where $\n$ is the outward normal unit vector to $\Omega$. On the exit site $\p\Omega_{A_2}$, we impose constant ionic densities,
\begin{equation}\label{eq:bc_2}
c_p(\x,t) = C_0\,, \quad c_m(\x,t) = C_0\,, \quad \x \in \p\Omega_{A_2}\,,
\end{equation}
and the remaining boundary part is impermeable to the flow of ions,
\begin{equation}\label{eq:bc_r}
\n \cdot j_p(\x,t) = 0\,, \quad \n \cdot j_m(\x,t) = 0\,, \quad \x \in \p\Omega_r\,.
\end{equation}
The voltage $v(\x,t)$ within the domain $\Omega$ is then the solution of Poisson's equation
\begin{equation}\label{eq:pois}
\Delta v(\x,t) + \frac{\FF}{\eps\eps_0}\left(c_p(\x,t) - c_m(\x,t) \right) = 0\,, \quad \x \in \Omega\,,
\end{equation}
where $\FF$ is the Faraday constant, while $\eps$ and $\eps_0$ are the relative and vacuum permittivities (see Table~\ref{table:param} for parameter values). Finally, we neglect here the capacitance of the membrane and impose that the component of the electric field normal to the boundary $\p\Omega$ is zero everywhere except for the narrow exit windows, where the voltage is grounded,
\begin{align}
\n \cdot \nabla v(\x,t) &= 0\,, \quad \x \in \p\Omega\backslash\p\Omega_{A_2}, \label{eq:noflux_bc} \\
v(\x,t) &= 0\,, \quad \x \in \p\Omega_{A_2}\,. \label{eq:ground}
\end{align}
We discuss a possible alternative boundary condition that treats the membrane as a capacitance in the Appendices; see also \cite{cartailler2018neuron}.
\section{Steady-state voltage distribution and voltage-current relation}
\begin{figure*}[!ht]
\begin{center}
\includegraphics[width=\linewidth]{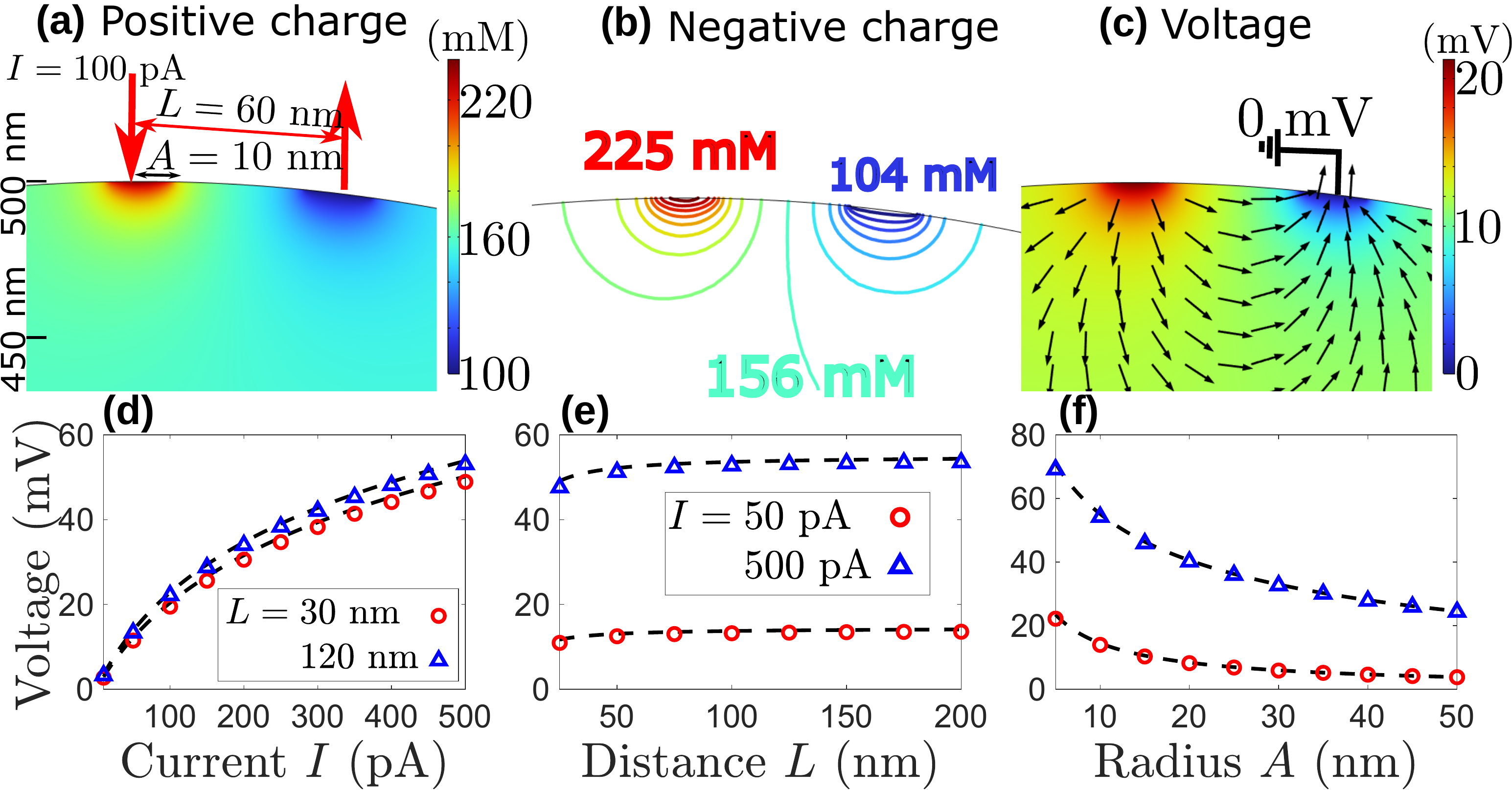}
\caption{\label{fig:I_V_curves} \textbf{Voltage nanodomain and I--V relation.} (\textbf{a})--(\textbf{c}) Voltage nanodomain near two channels with $I=100$ pA, $L=60$ nm and $A=10$ nm on a spherical domain of radius $R=500$ nm. (\textbf{d})--(\textbf{f}) Voltage vs current $I$, distance $L$ and radius $A$, with symbols (circles and triangles) indicating COMSOL numerical solutions, while dashed curves are calculated using formula \eqref{eq:vdiff} with $F(A_1,A_2)$ evaluated using Eq.~\eqref{eq:F_sph}. For other electro-diffusion model parameters; see Table \ref{table:param}.}
\end{center}
\end{figure*}

At steady-state, the concentrations $c_p(\x)$ and $c_m(\x)$, and voltage $v(\x)$ are solutions of the conservation laws $\nabla \cdot j_p(\x) = 0$ and $\nabla \cdot j_m(\x) = 0$ coupled to the Poisson's Eq.~\eqref{eq:pois}. In response to the influx through $\p\Omega_{A_1}$, negative charges accumulate at the influx channel region to compensate the local excess of positive charges. Using then local electro-neutrality as an ansatz yields the following approximation for the ionic concentrations at the influx location $\x_1$
\begin{equation}\label{eq:cdiff}
c_p(\x_1) = c_m(\x_1) \approx C_0 + \frac{IF\left(A_1,A_2\right)}{2\pi\FF D_p A_1}\,,
\end{equation}
with
\begin{widetext}
\begin{equation}\label{eq:F_gen}
\begin{split}
F(A_1,A_2) &= 1 + \frac{\pi A_1}{4A_2} - \frac{A_1}{4}\left(H(\x_1)\left(\log\left(\frac{A_1}{R}\right) - \frac{1}{2} \right) + H(\x_2)\left(\log\left(\frac{A_2}{R}\right) - 1 + \log(2)\right)\right) \\
&+ \frac{\pi A_1}{R}\left(\RR(\x_1;\x_1) + \RR(\x_2;\x_2) - 2 G(\x_1;\x_2)\right) + O\left(\frac{A_1^2}{R^2} \right) + O\left(\frac{A_1A_2}{R^2} \right)\,,
\end{split}
\end{equation}
\end{widetext}
where $A_1$ and $A_2$ are the narrow channels radii. Note that we introduce a characteristic length-scale $R$, with $A_1,\,A_2 \ll R$, which represents the radius when $\Omega$ is a ball. For convex domains with an isoperimetric ratio satisfying $|\p\Omega|/|\Omega|^{2/3} \sim O(1)$, we define the length-scale $R$ as the radius of the largest ball $B(\x,r)$ contained within $\Omega$ and tangent to the boundary,
\begin{equation}\label{eq:def_R}
R = \max_{r}\left\{ r \, | B(\x,r) \, \in \Omega \right\}\,.
\end{equation}
In Eq.~\eqref{eq:F_gen}, $H(\x_1)$ and $H(\x_2)$ are the mean surface curvature functions at the center of each channel, while $\RR(\x;\y)$ is the regular part of $G(\x;\y)$, the Neumann Green's function on $\Omega$ solution of (see Appendices)
\begin{subequations}
\begin{align}
\Delta G(\x;\y) &= \frac{R}{|\Omega|}\,, \quad \x \in \Omega\,, \y \in \p\Omega\,, \\
\n \cdot \nabla G(\x;\y) &= R\delta(\x-\y)\,, \quad \x,\y \in \p\Omega\,.
\end{align}
\end{subequations}
We compute the voltage from the negative charge concentration by using the Boltzmann's solution,
\begin{equation}
v(\x) = \frac{k_B\TT}{e}\log\left(\frac{C_m(\x)}{C_0}\right)\,,
\end{equation}
and after substituting Eq.~\eqref{eq:cdiff}, we get
\begin{equation}\label{eq:vdiff}
v(\x_1) = \frac{k_B\TT}{e}\log\left(1 + \frac{IF\left(A_1,A_2\right)}{2\pi\FF D_pC_0A_1}\right)\,.
\end{equation}
To explore how the current $I$, the distance $L$ and the two channel radii $A_1$ and $A_2$ affect the voltage difference generated by a positive ions influx, we consider a ball of radius $R$. We simulate with COMSOL Multiphysics \cite{comsol} the influx and efflux from two identical channels of common radius $A$ [Fig.~\ref{fig:I_V_curves}(\textbf{a})-(\textbf{c})]. We find nonlinear voltage effects at high current amplitude for $I > 100$ pA [Fig.~\ref{fig:I_V_curves}(\textbf{d})], while significant changes due to the distance $L$ only happen when the channels are nearby [Fig.~\ref{fig:I_V_curves}(\textbf{e})]. These results are in agreement with Eq.~\eqref{eq:vdiff}, where the voltage evolves as the reciprocal of the channel radius $A$ [Fig.~\ref{fig:I_V_curves}(\textbf{f})]. \\
\section{Voltage changes measured by the penetration length}
\begin{figure*}[!ht]
\begin{center}
\includegraphics[width=\linewidth]{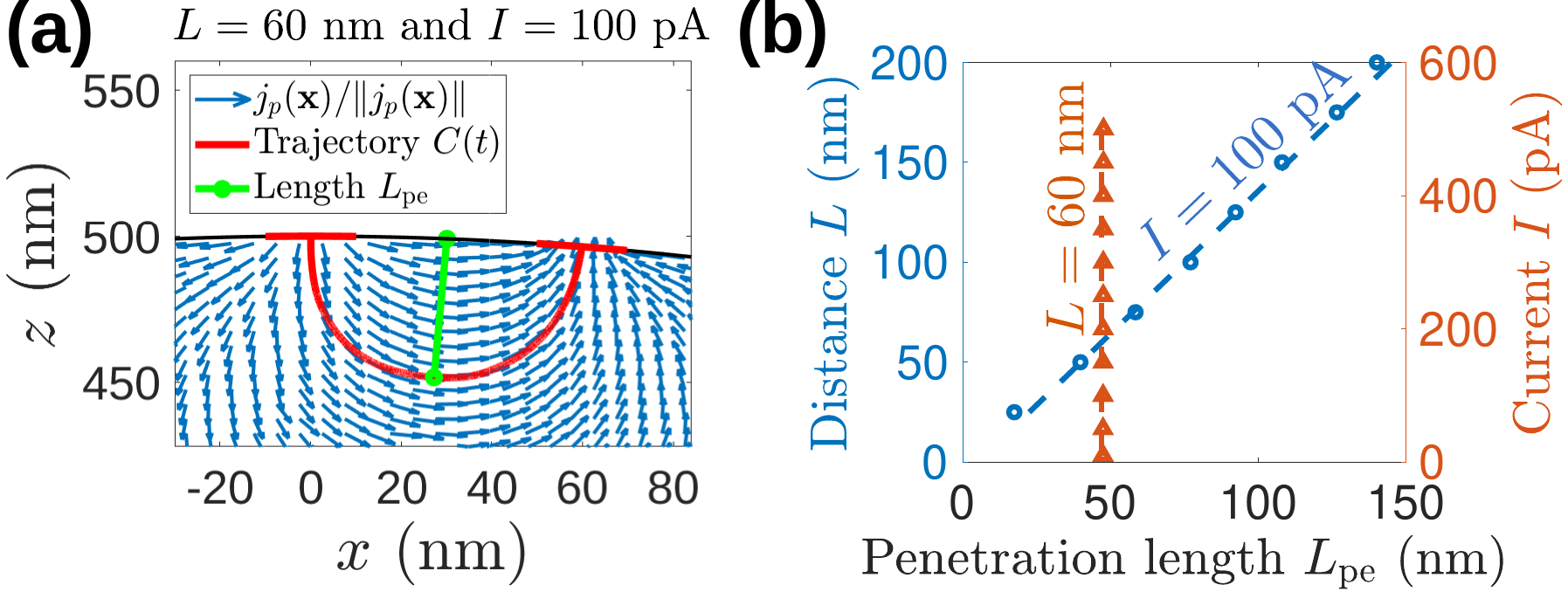}
\caption{\label{fig:Lpe}{\textbf{Voltage penetration length.} (\textbf{a}) The penetration length $L_\text{pe}$ (green line) is the deepest point along the trajectory $C(t)$ solution of Eq.~\eqref{eq:dyna} that connects the two channels. (\textbf{b}) Penetration length $L_\text{pe}$ vs current $I$ and distance $L$ compared to a numerical fit (dashed lines). Here, the windows have a common radius of $A=10$ nm, with domain and other parameters the same as in Fig.~\ref{fig:I_V_curves}.}}
\end{center}
\end{figure*}

To estimate the size of voltage nanodomains, we use the penetration length $L_\text{pe}$ defined as the deepest point along the fluxline $C(t)$ starting at the center of the influx channel $\p\Omega_{A_1}$ and ending in the neighboring target $\p\Omega_{A_2}$, representing an exchanger or an ionic channel. This trajectory is tangent to the gradient flow $j_p(\x)$, thus the solution of
\begin{equation}\label{eq:dyna}
\frac{dC(t)}{dt} = j_p\left(C(t)\right),
\end{equation}
with $C(0) = \x_1$. The penetration length $L_\text{pe}$ is the deepest point along $C(t)$ from the local boundary (Fig.~\ref{fig:Lpe}(\textbf{a})),
\begin{equation}
L_\text{pe} = \max_t\min_{\x \in \p\Omega}\left\{ \left. \|\x - C(t)\| \, \right| \, C(t) \in \Omega \right\}\,.
\end{equation}
We report here in the case of a ball that the penetration length can vary from tens to hundreds of nanometers [Fig.~\ref{fig:Lpe}(\textbf{b})], depending on the Euclidean distance $L$ between the channels. Note also that the penetration length is unaffected by the current magnitude $I$. These results were obtained for each parameter set by numerically solving Eq.~\eqref{eq:dyna} with the simulated gradient flows from Fig.~\ref{fig:I_V_curves}(\textbf{d})-(\textbf{e}). We then use the ansatz $L_\text{pe} = \beta_1 + \beta_2 L$ to fit the numerical results [dashed lines in Fig.~\ref{fig:Lpe}(\textbf{b})]. We find $\beta_1 \approx 5$ nm and $\beta_2 \approx 0.7$ nm$^{-2}$ and conclude that the penetration length is of the same order of magnitude as the distance $L$. We expect similar results for more complicated geometries than spherical domains when the narrow channels are located within a close neighborhood. \\
\section{Membrane curvature modulates sub-cellular voltage}
\begin{figure*}[!ht]
\begin{center}
\includegraphics[width=\linewidth]{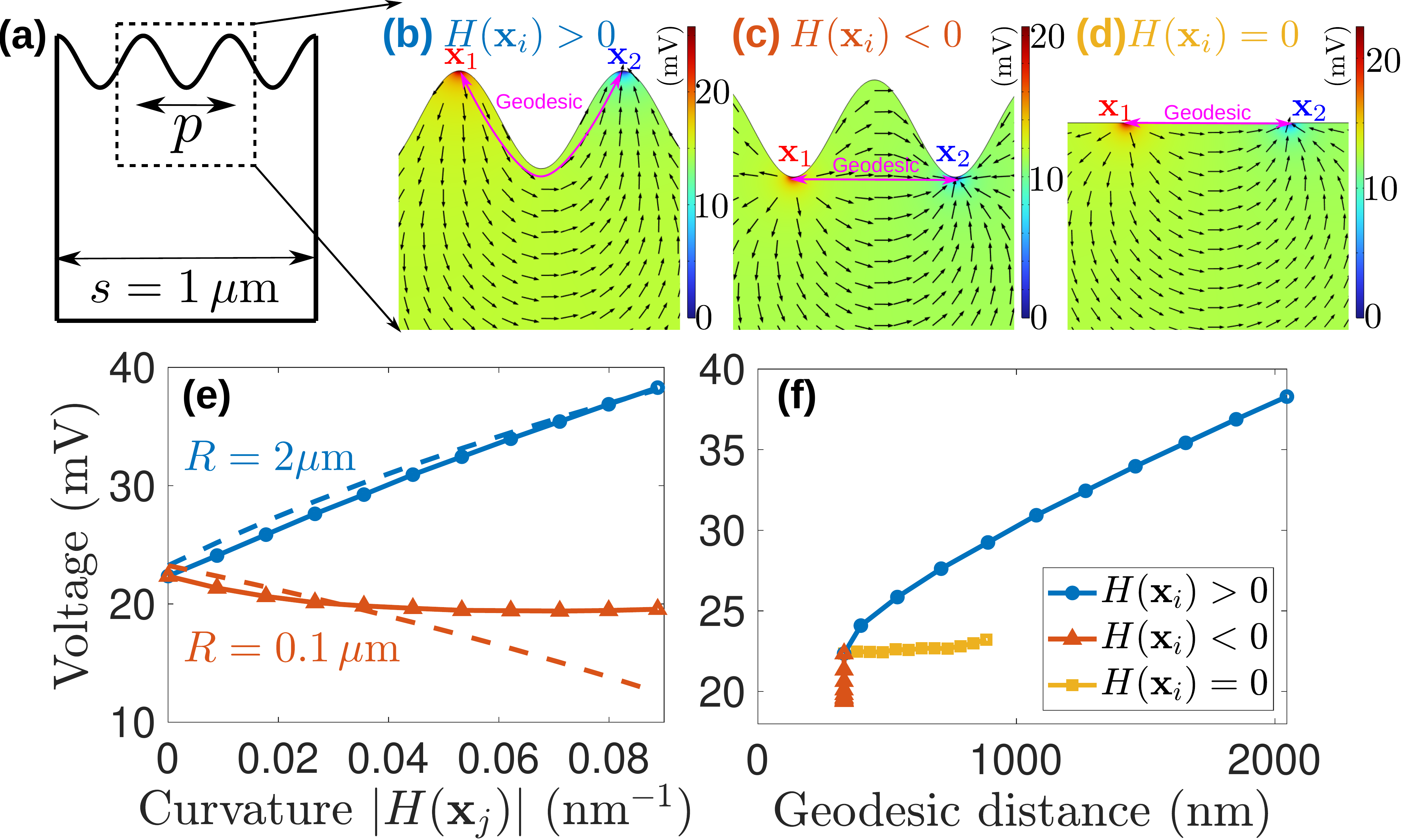}
\caption{\label{fig:curvature} \textbf{Membrane curvature modulates voltage dynamics.} (\textbf{a}) Lateral view of the hexahedral domain $\Omega$ with edges of length $s$ and oscillating upper boundary. (\textbf{b})--(\textbf{d}) Zoom on voltage nano-regions near channels located at points of positive, negative and zero mean curvature. (\textbf{e})--(\textbf{f}) Voltage vs mean curvature and shortest (geodesic) distance. The dashed curves in (\textbf{e}) are evaluated using the truncated voltage formula Eq.~\eqref{eq:vdiff_truncated} with $R=2\,\mu$m (blue dashed curve) and $R=0.1\,\mu$m (red dashed curve). Parameters are $s=1\,\mu$m, $I = 100$ pA, and $A = 10$ nm, with others taken from Table~\ref{table:param}.}
\end{center}
\end{figure*}

It has been suggested \cite{cartailler2017jns} that the membrane curvature can locally modulate the voltage, but the exact influence remains unclear. Thus, to quantify the role of curvature, we study a generic hexahedral domain $\Omega$ with edges of length $s$ and an oscillating upper boundary [Fig.~\ref{fig:curvature}(\textbf{a})] parameterized as
\begin{equation}\label{eq:osc_surf}
\gamma_u(x,y) = \left(x,y,s + \gamma(x)\right), \quad 0 \leq x,y \leq s\,,
\end{equation}
where $\gamma(x) =  k\cos\left(\frac{2\pi x}{p}\right)$ has an amplitude of $k$ and a period of $p=s/3$. The mean surface curvature $H(\x)$ evaluated on the upper boundary is given analytically by \cite{struik1988},
\begin{equation}
H(\x) = \frac{-\gamma''(x)}{2\left(1+\left(\gamma'(x)\right)^2\right)^{3/2}}\,,
\end{equation}
leading to
\begin{equation}
H(\x) = \frac{\frac{2\pi^2}{p^2}k\cos\left(\frac{2\pi}{p}x\right)}{\left(1 + \frac{4\pi^2 k^2}{p^2} \sin^2\left(\frac{2\pi}{p}x\right) \right)^{3/2}}\,.
\end{equation}
We then add two narrow circular windows of common radius $A$ to the upper boundary section, around $\x_1 = \gamma_u(s/3,s/2)$ and $\x_2 = \gamma_u(2s/3,s/2)$ [Fig.~\ref{fig:curvature}(\textbf{a})], such that the channels are located one period apart where the curvature is maximal and equal to
\begin{equation}
H(\x_1) = H(\x_2) = \frac{2\pi^2 k}{p^2}\,.
\end{equation}
The amplitude $k$ is then used to control the domain deformation and the mean curvature sign, with $H(\x_i)$ becoming positive [Fig.~\ref{fig:curvature}(\textbf{b})] when $k>0$ or negative when $k<0$ [Fig.~\ref{fig:curvature}(\textbf{c})]. Interestingly, while increasing a positive curvature leads to higher voltages [blue curve in Fig.~\ref{fig:curvature}(\textbf{e})], the opposite trend occurs for a negative curvature [red curve in Fig.~\ref{fig:curvature}(\textbf{e})], where ionic flow from $\p\Omega_{A_1}$ to $\p\Omega_{A_2}$ is facilitated by the membrane shape. This can be seen by analyzing the shortest (or geodesic) path between the channels: when $H(\x_i)>0$ the path is parallel to $\p\Omega$ and its length is calculated as
\begin{equation}
d(\x_1,\x_2) = \int_{x_1}^{x_2} \sqrt{1 + \left( \gamma'(x) \right)^2} dx\,,
\end{equation}
which increases due to boundary changes [blue curve in Fig.~\ref{fig:curvature}(\textbf{f})], while when $H(\x_i) < 0$ the shortest path avoids the boundary and remains unaffected by curvature variations [red vertical line in Fig.~\ref{fig:curvature}(\textbf{f})]. We compare our simulation results against Eq.~\eqref{eq:vdiff} truncated as
\begin{equation}\label{eq:vdiff_truncated}
\begin{split}
v(\x_1) &\approx \frac{k_B\TT}{e}\log\left(1 + \frac{I}{2\pi\FF D_p C_0 A}\left( 1 + \frac{\pi}{4} \right. \right. \\
& \left. \left. + \left(H(\x_1) + H(\x_2)\right)\frac{A}{4}\log\left(\frac{R}{A}\right) \right)\right)\,,
\end{split}
\end{equation}
which allows us to recover the appropriate voltage growth ($H(\x_1),\,H(\x_2) > 0$), or decay ($H(\x_1),\,H(\x_2) < 0$). Here an estimate of the characteristic length-scale $R$ is required. However, the definition given earlier in Eq.~\eqref{eq:def_R} does not necessarily hold due to the non-convexity of $\Omega$ and its large boundary curvature variations. Thus, we decided to use numerical simulations to adjust the slope of Eq.~\eqref{eq:vdiff_truncated} resulting in optimal $R$ values: $R = 2\,\mu$m in the positive curvature case [blue dashed curve in Fig.~\ref{fig:curvature}(\textbf{e})], and $R = 0.1\,\mu$m for its negative counterpart [red dashed curve in Fig.~\ref{fig:curvature}(\textbf{e})], where only the initial decay before the flattening of the voltage curve was fitted (curvature changes beyond a certain threshold have no influence). Note that how to define characteristic length-scales in non-convex domains remains an open problem. Finally we conclude with the case of a flat boundary [Fig.~\ref{fig:curvature}(\textbf{d})], where we find that, in comparison to curvature changes, increasing the distance between channels leads to almost no voltage variations [Fig.~\ref{fig:curvature}(\textbf{f})].
\section{Voltage distribution in dendritic spines}
\begin{figure*}[!ht]
\begin{center}
\includegraphics[width=\linewidth]{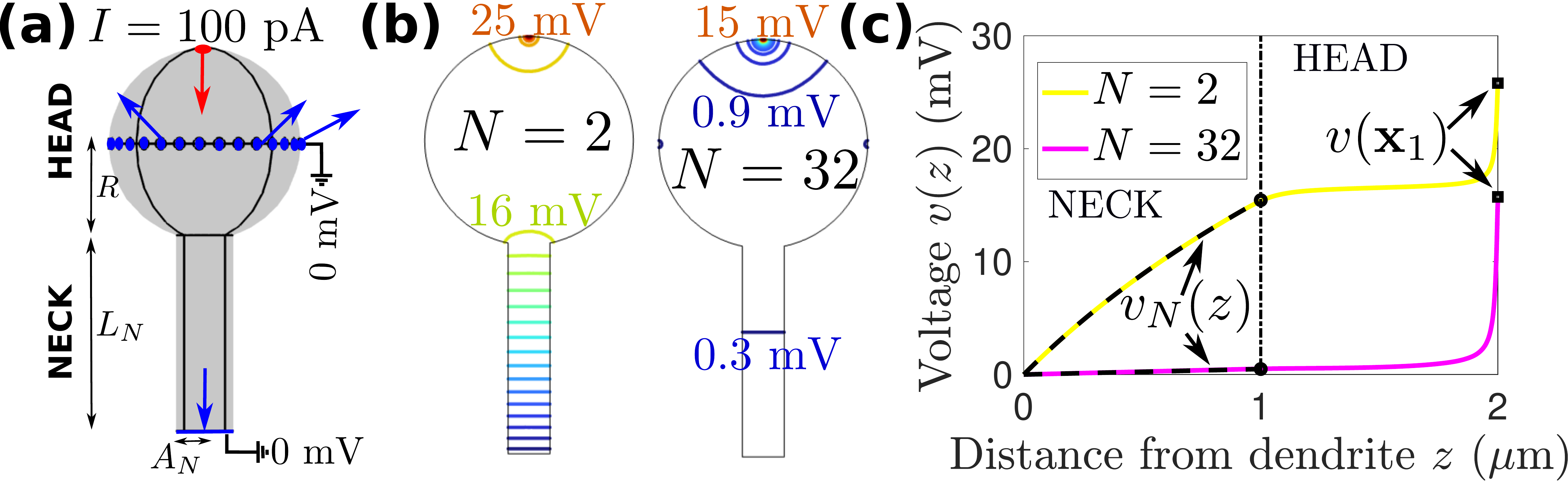}
\caption{\label{fig:shunt} \textbf{Voltage distribution in a dendritic spine.} (\textbf{a}) Dendritic spine modeled as a ball connected to a narrow cylindrical neck, with additional $(N-2)$ exit channels facilitating ionic flow. (\textbf{b}) Equipotential lines without (left) and with (right) additional exit channels on the spherical head. (\textbf{c}) Voltage vs distance from dendrite. Color curves correspond to simulation results while the black dashed neck solutions are evaluated using Eq.~\eqref{eq:vdiff_n}. Parameters: $R = 0.5\,\mu\text{m}$, $L_N = 1\,\mu\text{m}$, $A_n=0.01\,\mu\text{m}$ $(n\neq N)$, and $A_N = 0.1\,\mu\text{m}$, with others from Table~\ref{table:param}.}
\end{center}
\end{figure*}

The voltage distribution in a dendritic spine is a key regulator of synaptic transmission. Yet how a synaptic current is converted into a voltage map is still unclear due to the complex geometrical nano-organization. Such conversion is controlled by the narrow neck passage \cite{cartailler2018neuron,doyon2020} which limits ionic motion to the parent dendrite \cite{yuste2010dendritic,biess2007diffusion}. Here, we report an unexpected phenomenon where the presence of conducting channels on the spine head surface facilitates a direct efflux of ions and, thus, modifies the voltage distribution in the entire spine. To show this effect, we consider a spine geometry consisting of a ball of radius $R$ connected to an external narrow window $\p\Omega_{A_N}$ by a thin cylinder of length $L_N$ and radius $A_N$ [Fig.~\ref{fig:shunt}(\textbf{a})], modeling the narrow neck \cite{yuste2010dendritic}. We add multiple conducting channels $\p\Omega_{A_n}$ with $n=2,\ldots,N-1$ on the head that facilitate the efflux of ions. Window $\p\Omega_{A_1}$ receives a positive charge influx [Eq.~\eqref{eq:influx}], while on the exit sites constant ionic densities and voltage ground conditions are imposed
\begin{equation}
c_p(\x) = C_0,\, c_m(\x) = C_0,\, v(\x) = 0,\, \quad \x \in \p\Omega_{A_n}
\end{equation}
for $n=2,\ldots,N$. The total boundary flux exiting through the upper head section $\Phi_H$ or via the neck $\Phi_N$ is given by
\begin{equation}
\Phi_H = \sum_{n=2}^{N-1}\int_{\p\Omega_{A_n}} \hspace{-7mm} \n \cdot j_p(\x) d\x, \quad \Phi_N = \int_{\p\Omega_{A_N}} \hspace{-7mm} \n \cdot j_p(\x) d\x\,,
\end{equation}
and the flux ratio $P_N = \Phi_N/(\Phi_N + \Phi_H)$ exiting through the neck depends on the geometry, as revealed by the explicit formula,
\begin{equation}\label{eq:P_N}
P_N = \frac{1}{1+\frac{4}{\pi}\left(1+\frac{L_N}{A_N}\right)\sum_{n=2}^{N-1} \frac{A_n}{A_N}}\,.
\end{equation}
At the source location [square symbols in Fig.~\ref{fig:shunt}(\textbf{c})] the voltage is calculated as
\begin{equation}\label{eq:vdiff_ds}
v(\x_1) = \frac{k_B\TT}{e}\log\left(1 + \frac{I\left(\left(L_N + A_N\right)P_N \hspace{-1mm} + \hspace{-1mm} \frac{A_N^2}{A_1}\right)}{2\pi\FF D_p C_0 A_N^2}\hspace{-1mm}\right)\hspace{-1mm},
\end{equation}
after which it decreases quickly within the spine head, where far away from the channels it is almost equipotential [Fig.~\ref{fig:shunt}(\textbf{b})]. The neck voltage drop is then
\begin{equation}\label{eq:vdiff_n}
v_N(z) = \frac{k_B\TT}{e}\log\left(1 + \frac{Iz}{2\FF D_p C_0 \pi A_N^2} P_N\right)\,.
\end{equation}
Interestingly by comparing two distributions with multiple exit channels $(N=32)$ and without $(N=2)$ [Fig.~\ref{fig:shunt}(\textbf{b})--(\textbf{c})], we find that for $N$ large the probability ratio $P_N$ [Eq.~\eqref{eq:P_N}] nearly vanishes and the voltage becomes zero everywhere except for the neighborhood of the source, yielding the expression
\begin{equation}
v(\x_1) \approx \frac{k_B\TT}{e}\log\left(1 + \frac{I}{2\FF D_p C_0 \pi A_1}\right)
\end{equation}
with $v_N(z) \approx 0$. We conclude that the neck geometry, therefore, ceases to matter, with the influx window radius left as the only geometrical parameter controlling the current conversion into voltage, but with the voltage that still decays from the head to the base of the dendritic spine.
\section{Discussion and perspectives}
Using a two-charge electro-diffusion model based on the Poisson--Nernst--Planck equations, we study here the voltage penetration in nanoregions due to ionic fluxes from membrane channels, the role of which is to modulate nearby channels \cite{zamponi2011}. We obtain explicit current--voltage relations [Fig.~\ref{fig:I_V_curves}], highlighting the role of the size of the channels, their organization, and the membrane curvature. Furthermore, we quantify the penetration of an influx current within a cellular electrolyte using a length-scale based on the geometry of the fluxline connecting the channels [Fig.~\ref{fig:Lpe}]. \\
We also report how boundary deformations can affect current attenuation, with voltage either increasing or decreasing depending on the mean curvature sign at the boundary flux locations [Fig.~\ref{fig:curvature}]. Finally, motivated by the specific morphology of dendritic spines, we predict that the voltage conversion due to multiple channel windows on the head can shunt the attenuation due to the narrow neck connecting the head to the bulk compartment [Fig.~\ref{fig:shunt}]. The present predictions could be extended to transient dynamics following a time-dependent ionic influx.
\section{Acknowledgements} 
F.~P.-L.~gratefully acknowledges the support from the Natural Sciences and Engineering Research Council of Canada (NSERC) (Award No.~578183-2023) and from the Fondation ARC (Award No.~ARCPDF12020020001505) through postdoctoral fellowships. D.~H.~was supported by the European Research Council (ERC) under the European Union’s Horizon 2020 Research and Innovation Program (grant agreement No.~882673) and ANR AstroXcite.
\begin{appendix}
\section{Explicit Neumann Green's function}
We first recall the voltage formula Eq.~\eqref{eq:vdiff} for two narrow channels $\p\Omega_{A_1}$ and $\p\Omega_{A_2}$ on an otherwise reflective membrane $\p\Omega$,
\begin{equation}
v(\x_1) = \frac{k_B\TT}{e}\log\left(1 + \frac{IF\left(A_1,A_2\right)}{2\pi\FF D_pC_0A_1}\right)\,,
\end{equation}
where $F(A_1,A_2)$ is a function of the small radii $A_1$ and $A_2$ that is derived using the Neumann Green's function $G(\x;\y)$. On a domain $\Omega$ of length-scale $R$, of the order $R\sim O(\sqrt{|\p\Omega|})$ (which is well-defined by the local curvature radius for domains that are convex and of order one isoperimetric ratio $|\p\Omega|/|\Omega|^{2/3} \sim O(1)$), the Neumann Green's function is the solution of
\begin{widetext}
\begin{equation}\label{eq:gf}
\Delta G(\x;\y) = \frac{R}{|\Omega|}\,, \quad \x \in \Omega\,, \y \in \p\Omega\,, \quad \n \cdot \nabla G(\x;\y) = R\delta(\x-\y)\,, \quad \x,\y \in \p\Omega \\
\end{equation}
with the constraint $\int_\Omega G(\x;\y)d\x = 0$. Near the singular diagonal with $\|\x - \y\| \ll R$, the Green's function can be expanded as
\begin{equation}
G(\x;\y) \approx \frac{R}{2\pi\|\x-\y\|} - \frac{RH(\y)}{4\pi}\log\left(\frac{\|\x-\y\|}{R}\right) + \RR(\x;\y)\,,
\end{equation}
where $H(\y)$ is the mean surface curvature evaluated at $\y$ on $\p\Omega$, and the function $\RR(\x;\y)$ is regular. In the narrow target approximation, i.e., with $A_1,A_2 \ll R$, the function $F(A_1,A_2)$ can be expressed as [see Eq.~\eqref{eq:F_gen}]
\begin{equation}\label{eq:F_gen_app}
\begin{split}
F(A_1,A_2) &= 1 + \frac{\pi A_1}{4A_2} - \frac{A_1}{4}\left(H(\x_1)\left(\log\left(\frac{A_1}{R}\right) - \frac{1}{2} \right) + H(\x_2)\left(\log\left(\frac{A_2}{R}\right) - 1 + \log(2)\right)\right) \\
& + \frac{\pi A_1}{R}\left(\RR(\x_1;\x_1) + \RR(\x_2;\x_2) - 2 G(\x_1;\x_2)\right) + O\left(\frac{A_1^2}{R^2} \right) + O\left(\frac{A_1A_2}{R^2} \right)\,,
\end{split}
\end{equation}
which can be truncated as
\begin{equation}\label{eq:F_truncated_app}
F(A_1,A_2) = 1 + \frac{\pi A_1}{4A_2} - \frac{A_1}{4}\left(H(\x_1)\log\left(\frac{A_1}{R}\right) + H(\x_2)\log\left(\frac{A_2}{R}\right)\right) + O\left(\frac{A_1}{R}\right)\,,
\end{equation}
when the exact Neumann Green's function is unknown. Exact solutions for $G(\x;\y)$ are only available on certain specific geometries, such as the ball of radius $R$ for which we have \cite{cheviakov2010},
\begin{equation}\label{eq:gf_sphere}
G(\x;\y) =\frac{R}{2\pi\|\x-\y\|} - \frac{1}{4\pi}\log\left(\frac{\|\x-\y\|^2}{2R^2} +\frac{\|\x-\y\|}{R}\right) + \RR(\x;\y)\,, \quad \|\x\| = \|\y\| = R \,, \quad \x \neq \y
\end{equation}
with the regular part $\RR(\x;\y) = \left(\log(2) - \frac{9}{5}\right)/(4\pi)$ that is constant everywhere on the boundary. Finally when substituting these expressions within Eq.~\eqref{eq:F_gen_app}, and using $H(\x) = 1/R$ for the mean curvature function, we obtain the asymptotic expansion as follows:
\begin{equation}\label{eq:F_sph}
F(A_1,A_2) = 1 + \frac{\pi A_1}{4A_2} - \frac{A_1}{4R}\log\left(\frac{A_1A_2}{R^2}\right) + \frac{A_1}{R}\left(\frac{3}{8} - \frac{\log(2)}{4} - \frac{R}{L} + \frac{1}{2}\log\left( \frac{L^2}{2R^2} + \frac{L}{R}\right)\right) + O\left(\frac{A_1^2}{R^2} \right) + O\left(\frac{A_1A_2}{R^2} \right)\,,
\end{equation}
where $L=\| \x_1 - \x_2 \|$ denotes the Euclidean distance between two narrow channels on a spherical domain.
\end{widetext}
\section{Membrane capacitance boundary condition}
An alternative to the voltage no-flux boundary condition [Eq.~\eqref{eq:noflux_bc}] is to treat the membrane as a capacitance. In that case, the voltage changes across the membrane proportionally to the normal component of the electric field, leading to the equation \cite{cartailler2018neuron,doyon2020}
\begin{equation}\label{eq:robin}
\n \cdot \nabla v(\x) = \frac{\eps_{\text{mem}}}{\eps} \frac{v_{\text{ext}} - v(\x)}{d}\,, \quad \x \in \p\Omega\backslash\p\Omega_{A_2}\,,
\end{equation}
where $\eps_{\text{mem}}$ and $\eps$ are the membrane and water relative permittivities, $d$ is the width of the membrane and $v_{\text{ext}}$ is the constant external voltage. Thus, neglecting the membrane capacitance is equivalent to setting the relative membrane permittivity to zero, even though for typical membranes we have $\eps_{\text{mem}} = 10$. Next, the coefficient $\kappa = \eps_{\text{mem}}R/(\eps d)$ measures the coupling strength in Eq.~\eqref{eq:robin} in comparison with the domain length-scale $R$, and it evaluates to $\kappa \approx 9$ using $d=7$ nm, $R=500$ nm, and $\eps = 78.4$. However, simulation results show that the current conversion into voltage is barely affected by the introduction of a membrane capacitance \cite{cartailler2018neuron}, which justifies the simplified no-flux condition [Eq.~\eqref{eq:noflux_bc}] and allows us to derive current--voltage (I--V) relations.
\section{Model parameters}
We summarize in Table~\ref{table:param} the parameter values used for the simulations.
\begin{center}
\begin{table}[!ht]
\centering
\caption{\label{table:param} \textbf{Model parameters and physical constants.}}
\begin{tabular}{|l|c|l|}
\hline
\textbf{Parameter} & \textbf{Symbol} & \textbf{Value} \\
\hline
Domain length-scale & $R$ & $500$ nm \\
Number of channels & $N$ & $\geq 2$ \\
Channel radii & $A_1,\ldots,A_{N-1}$ & $10$ nm \\
Distance & $L = \|\x_1-\x_2\|$ & $[0.025,\,1]$ $\mu$m \\
Neck radius & $A_N$ & $100$ nm \\
Neck length & $L_N$ & $1\,\mu$m \\
Diffusion coefficients & $D_p,D_m$ & $200\,\mu$m$^2$s$^{-1}$ \\
Concentration & $C_0$ & $100$ mM \\
Current amplitude & $I$ & $[10,\,500]$ pA \\
Electron charge & $e$ & $1.60 \times 10^{-19}$ C \\
Avogadro constant & $N_A$ & $6.02 \times 10^{23}$ mol$^{-1}$ \\
Boltzmann constant & $k_B$ & $1.38 \times 10^{-23}$ J K$^{-1}$ \\
Temperature & $\mathcal{T}$ & $298$ K \\
Relative permittivity & $\eps$ & $78.4$ \\
Vacuum permittivity & $\eps_0$ & $8.85\times 10^{-12}$ C V$^{-1}$m$^{-1}$ \\
Faraday constant & $\FF$ & $96\,485.33$ C mol$^{-1}$ \\
\hline
\end{tabular}
\end{table}
\end{center}
\end{appendix}
\normalem

\begin{thebibliography}{32}%
\makeatletter
\providecommand \@ifxundefined [1]{%
 \@ifx{#1\undefined}
}%
\providecommand \@ifnum [1]{%
 \ifnum #1\expandafter \@firstoftwo
 \else \expandafter \@secondoftwo
 \fi
}%
\providecommand \@ifx [1]{%
 \ifx #1\expandafter \@firstoftwo
 \else \expandafter \@secondoftwo
 \fi
}%
\providecommand \natexlab [1]{#1}%
\providecommand \enquote  [1]{``#1''}%
\providecommand \bibnamefont  [1]{#1}%
\providecommand \bibfnamefont [1]{#1}%
\providecommand \citenamefont [1]{#1}%
\providecommand \href@noop [0]{\@secondoftwo}%
\providecommand \href [0]{\begingroup \@sanitize@url \@href}%
\providecommand \@href[1]{\@@startlink{#1}\@@href}%
\providecommand \@@href[1]{\endgroup#1\@@endlink}%
\providecommand \@sanitize@url [0]{\catcode `\\12\catcode `\$12\catcode
  `\&12\catcode `\#12\catcode `\^12\catcode `\_12\catcode `\%12\relax}%
\providecommand \@@startlink[1]{}%
\providecommand \@@endlink[0]{}%
\providecommand \url  [0]{\begingroup\@sanitize@url \@url }%
\providecommand \@url [1]{\endgroup\@href {#1}{\urlprefix }}%
\providecommand \urlprefix  [0]{URL }%
\providecommand \Eprint [0]{\href }%
\providecommand \doibase [0]{https://doi.org/}%
\providecommand \selectlanguage [0]{\@gobble}%
\providecommand \bibinfo  [0]{\@secondoftwo}%
\providecommand \bibfield  [0]{\@secondoftwo}%
\providecommand \translation [1]{[#1]}%
\providecommand \BibitemOpen [0]{}%
\providecommand \bibitemStop [0]{}%
\providecommand \bibitemNoStop [0]{.\EOS\space}%
\providecommand \EOS [0]{\spacefactor3000\relax}%
\providecommand \BibitemShut  [1]{\csname bibitem#1\endcsname}%
\let\auto@bib@innerbib\@empty
\bibitem [{\citenamefont {Hille}(2001)}]{hille}%
  \BibitemOpen
  \bibfield  {author} {\bibinfo {author} {\bibfnamefont {B.}~\bibnamefont
  {Hille}},\ }\href@noop {} {\emph {\bibinfo {title} {Ionic Channels of
  Excitable Membranes}}},\ \bibinfo {edition} {3rd}\ ed.\ (\bibinfo
  {publisher} {Sinauer, Sunderland},\ \bibinfo {year} {2001})\ p.\ \bibinfo
  {pages} {814}\BibitemShut {NoStop}%
\bibitem [{\citenamefont {Turrigiano}\ and\ \citenamefont
  {Nelson}(2004)}]{Turrigiano}%
  \BibitemOpen
  \bibfield  {author} {\bibinfo {author} {\bibfnamefont {G.~G.}\ \bibnamefont
  {Turrigiano}}\ and\ \bibinfo {author} {\bibfnamefont {S.~B.}\ \bibnamefont
  {Nelson}},\ }\href@noop {} {\bibfield  {journal} {\bibinfo  {journal} {Nat.
  Rev. Neurosci.}\ }\textbf {\bibinfo {volume} {5}},\ \bibinfo {pages} {97}
  (\bibinfo {year} {2004})}\BibitemShut {NoStop}%
\bibitem [{\citenamefont {Nicoll}(2017)}]{nicoll}%
  \BibitemOpen
  \bibfield  {author} {\bibinfo {author} {\bibfnamefont {R.~A.}\ \bibnamefont
  {Nicoll}},\ }\href@noop {} {\bibfield  {journal} {\bibinfo  {journal}
  {Neuron}\ }\textbf {\bibinfo {volume} {93}},\ \bibinfo {pages} {281}
  (\bibinfo {year} {2017})}\BibitemShut {NoStop}%
\bibitem [{\citenamefont {Huganir}\ and\ \citenamefont
  {Nicoll}(2013)}]{Huganir}%
  \BibitemOpen
  \bibfield  {author} {\bibinfo {author} {\bibfnamefont {R.}~\bibnamefont
  {Huganir}}\ and\ \bibinfo {author} {\bibfnamefont {R.}~\bibnamefont
  {Nicoll}},\ }\href@noop {} {\bibfield  {journal} {\bibinfo  {journal}
  {Neuron}\ }\textbf {\bibinfo {volume} {80}},\ \bibinfo {pages} {704}
  (\bibinfo {year} {2013})}\BibitemShut {NoStop}%
\bibitem [{\citenamefont {Yuste}(2010)}]{yuste2010dendritic}%
  \BibitemOpen
  \bibfield  {author} {\bibinfo {author} {\bibfnamefont {R.}~\bibnamefont
  {Yuste}},\ }\href@noop {} {\emph {\bibinfo {title} {Dendritic Spines}}}\
  (\bibinfo  {publisher} {The MIT Press},\ \bibinfo {year} {2010})\BibitemShut
  {NoStop}%
\bibitem [{\citenamefont {Bezanilla}(2008)}]{bezanilla2008}%
  \BibitemOpen
  \bibfield  {author} {\bibinfo {author} {\bibfnamefont {F.}~\bibnamefont
  {Bezanilla}},\ }\href@noop {} {\bibfield  {journal} {\bibinfo  {journal}
  {Nat. Rev. Mol. Cell. Biol.}\ }\textbf {\bibinfo {volume} {9}},\ \bibinfo
  {pages} {323} (\bibinfo {year} {2008})}\BibitemShut {NoStop}%
\bibitem [{\citenamefont {Simms}\ and\ \citenamefont
  {Zamponi}(2014)}]{zamponi2014}%
  \BibitemOpen
  \bibfield  {author} {\bibinfo {author} {\bibfnamefont {B.~A.}\ \bibnamefont
  {Simms}}\ and\ \bibinfo {author} {\bibfnamefont {G.~W.}\ \bibnamefont
  {Zamponi}},\ }\href@noop {} {\bibfield  {journal} {\bibinfo  {journal}
  {Neuron}\ }\textbf {\bibinfo {volume} {82}},\ \bibinfo {pages} {24} (\bibinfo
  {year} {2014})}\BibitemShut {NoStop}%
\bibitem [{\citenamefont {Turner}\ \emph {et~al.}(2011)\citenamefont {Turner},
  \citenamefont {Dustin},\ and\ \citenamefont {Zamponi}}]{zamponi2011}%
  \BibitemOpen
  \bibfield  {author} {\bibinfo {author} {\bibfnamefont {R.~W.}\ \bibnamefont
  {Turner}}, \bibinfo {author} {\bibfnamefont {A.}~\bibnamefont {Dustin}},\
  and\ \bibinfo {author} {\bibfnamefont {G.~W.}\ \bibnamefont {Zamponi}},\
  }\href@noop {} {\bibfield  {journal} {\bibinfo  {journal} {Channels}\
  }\textbf {\bibinfo {volume} {5}},\ \bibinfo {pages} {440} (\bibinfo {year}
  {2011})}\BibitemShut {NoStop}%
\bibitem [{\citenamefont {Liu}\ \emph {et~al.}(2022)\citenamefont {Liu},
  \citenamefont {Lu}, \citenamefont {Villette}, \citenamefont {Gou},
  \citenamefont {Colbert}, \citenamefont {Lai}, \citenamefont {Guan},
  \citenamefont {Land}, \citenamefont {Lee}, \citenamefont {Assefa},
  \citenamefont {Zollinger}, \citenamefont {Korympidou}, \citenamefont
  {Vlasits}, \citenamefont {Pang}, \citenamefont {Su}, \citenamefont {Cai},
  \citenamefont {Froudarakis}, \citenamefont {Zhou}, \citenamefont {Patel},
  \citenamefont {Smith}, \citenamefont {Ayon}, \citenamefont {Bizouard},
  \citenamefont {Bradley}, \citenamefont {Franke}, \citenamefont {Clandinin},
  \citenamefont {Giovannucci}, \citenamefont {Tolias}, \citenamefont {Reimer},
  \citenamefont {Dieudonné},\ and\ \citenamefont {St-Pierre}}]{Dieudonne}%
  \BibitemOpen
  \bibfield  {author} {\bibinfo {author} {\bibfnamefont {Z.}~\bibnamefont
  {Liu}}, \bibinfo {author} {\bibfnamefont {X.}~\bibnamefont {Lu}}, \bibinfo
  {author} {\bibfnamefont {V.}~\bibnamefont {Villette}}, \bibinfo {author}
  {\bibfnamefont {Y.}~\bibnamefont {Gou}}, \bibinfo {author} {\bibfnamefont
  {K.~L.}\ \bibnamefont {Colbert}}, \bibinfo {author} {\bibfnamefont
  {S.}~\bibnamefont {Lai}}, \bibinfo {author} {\bibfnamefont {S.}~\bibnamefont
  {Guan}}, \bibinfo {author} {\bibfnamefont {M.~A.}\ \bibnamefont {Land}},
  \bibinfo {author} {\bibfnamefont {J.}~\bibnamefont {Lee}}, \bibinfo {author}
  {\bibfnamefont {T.}~\bibnamefont {Assefa}}, \bibinfo {author} {\bibfnamefont
  {D.~R.}\ \bibnamefont {Zollinger}}, \bibinfo {author} {\bibfnamefont {M.~M.}\
  \bibnamefont {Korympidou}}, \bibinfo {author} {\bibfnamefont {A.~L.}\
  \bibnamefont {Vlasits}}, \bibinfo {author} {\bibfnamefont {M.~M.}\
  \bibnamefont {Pang}}, \bibinfo {author} {\bibfnamefont {S.}~\bibnamefont
  {Su}}, \bibinfo {author} {\bibfnamefont {C.}~\bibnamefont {Cai}}, \bibinfo
  {author} {\bibfnamefont {E.}~\bibnamefont {Froudarakis}}, \bibinfo {author}
  {\bibfnamefont {N.}~\bibnamefont {Zhou}}, \bibinfo {author} {\bibfnamefont
  {S.~S.}\ \bibnamefont {Patel}}, \bibinfo {author} {\bibfnamefont {C.~L.}\
  \bibnamefont {Smith}}, \bibinfo {author} {\bibfnamefont {A.}~\bibnamefont
  {Ayon}}, \bibinfo {author} {\bibfnamefont {P.}~\bibnamefont {Bizouard}},
  \bibinfo {author} {\bibfnamefont {J.}~\bibnamefont {Bradley}}, \bibinfo
  {author} {\bibfnamefont {K.}~\bibnamefont {Franke}}, \bibinfo {author}
  {\bibfnamefont {T.~R.}\ \bibnamefont {Clandinin}}, \bibinfo {author}
  {\bibfnamefont {A.}~\bibnamefont {Giovannucci}}, \bibinfo {author}
  {\bibfnamefont {A.~S.}\ \bibnamefont {Tolias}}, \bibinfo {author}
  {\bibfnamefont {J.}~\bibnamefont {Reimer}}, \bibinfo {author} {\bibfnamefont
  {S.}~\bibnamefont {Dieudonné}},\ and\ \bibinfo {author} {\bibfnamefont
  {F.}~\bibnamefont {St-Pierre}},\ }\href@noop {} {\bibfield  {journal}
  {\bibinfo  {journal} {Cell}\ }\textbf {\bibinfo {volume} {185}},\ \bibinfo
  {pages} {3408} (\bibinfo {year} {2022})}\BibitemShut {NoStop}%
\bibitem [{\citenamefont {Wolf}\ \emph {et~al.}(2019)\citenamefont {Wolf},
  \citenamefont {Segawa}, \citenamefont {Kondadi}, \citenamefont {Anand},
  \citenamefont {Bailey}, \citenamefont {Reichert}, \citenamefont {van~der
  Bliek}, \citenamefont {Shackelford}, \citenamefont {Liesa},\ and\
  \citenamefont {Shirihai}}]{emboJ}%
  \BibitemOpen
  \bibfield  {author} {\bibinfo {author} {\bibfnamefont {D.~M.}\ \bibnamefont
  {Wolf}}, \bibinfo {author} {\bibfnamefont {M.}~\bibnamefont {Segawa}},
  \bibinfo {author} {\bibfnamefont {K.}~\bibnamefont {Kondadi}}, \bibinfo
  {author} {\bibfnamefont {R.}~\bibnamefont {Anand}}, \bibinfo {author}
  {\bibfnamefont {S.~T.}\ \bibnamefont {Bailey}}, \bibinfo {author}
  {\bibfnamefont {A.~S.}\ \bibnamefont {Reichert}}, \bibinfo {author}
  {\bibfnamefont {A.~M.}\ \bibnamefont {van~der Bliek}}, \bibinfo {author}
  {\bibfnamefont {D.~B.}\ \bibnamefont {Shackelford}}, \bibinfo {author}
  {\bibfnamefont {M.}~\bibnamefont {Liesa}},\ and\ \bibinfo {author}
  {\bibfnamefont {O.~S.}\ \bibnamefont {Shirihai}},\ }\href@noop {} {\bibfield
  {journal} {\bibinfo  {journal} {EMBO J.}\ }\textbf {\bibinfo {volume} {38}},\
  \bibinfo {pages} {e101056} (\bibinfo {year} {2019})}\BibitemShut {NoStop}%
\bibitem [{\citenamefont {Lagache}\ \emph {et~al.}(2019)\citenamefont
  {Lagache}, \citenamefont {Jayant},\ and\ \citenamefont
  {Yuste}}]{Lagacheyuste}%
  \BibitemOpen
  \bibfield  {author} {\bibinfo {author} {\bibfnamefont {T.}~\bibnamefont
  {Lagache}}, \bibinfo {author} {\bibfnamefont {K.}~\bibnamefont {Jayant}},\
  and\ \bibinfo {author} {\bibfnamefont {R.}~\bibnamefont {Yuste}},\
  }\href@noop {} {\bibfield  {journal} {\bibinfo  {journal} {J Comput
  Neurosci}\ }\textbf {\bibinfo {volume} {47}},\ \bibinfo {pages} {77}
  (\bibinfo {year} {2019})}\BibitemShut {NoStop}%
\bibitem [{\citenamefont {Mc~Hugh}\ \emph {et~al.}(2023)\citenamefont
  {Mc~Hugh}, \citenamefont {Makarchuk}, \citenamefont {Mozheiko}, \citenamefont
  {Fernandez-Villegas}, \citenamefont {Kaminski~Schierle}, \citenamefont
  {Kaminski}, \citenamefont {Keyser}, \citenamefont {Holcman},\ and\
  \citenamefont {Rouach}}]{JeffHugh2023}%
  \BibitemOpen
  \bibfield  {author} {\bibinfo {author} {\bibfnamefont {J.}~\bibnamefont
  {Mc~Hugh}}, \bibinfo {author} {\bibfnamefont {S.}~\bibnamefont {Makarchuk}},
  \bibinfo {author} {\bibfnamefont {D.}~\bibnamefont {Mozheiko}}, \bibinfo
  {author} {\bibfnamefont {A.}~\bibnamefont {Fernandez-Villegas}}, \bibinfo
  {author} {\bibfnamefont {G.~S.}\ \bibnamefont {Kaminski~Schierle}}, \bibinfo
  {author} {\bibfnamefont {C.~F.}\ \bibnamefont {Kaminski}}, \bibinfo {author}
  {\bibfnamefont {U.~F.}\ \bibnamefont {Keyser}}, \bibinfo {author}
  {\bibfnamefont {D.}~\bibnamefont {Holcman}},\ and\ \bibinfo {author}
  {\bibfnamefont {N.}~\bibnamefont {Rouach}},\ }\href@noop {} {\bibfield
  {journal} {\bibinfo  {journal} {Nanoscale}\ }\textbf {\bibinfo {volume}
  {15}},\ \bibinfo {pages} {12245} (\bibinfo {year} {2023})}\BibitemShut
  {NoStop}%
\bibitem [{\citenamefont {Qian}\ and\ \citenamefont
  {Sejnowski}(1989)}]{qian1989}%
  \BibitemOpen
  \bibfield  {author} {\bibinfo {author} {\bibfnamefont {N.}~\bibnamefont
  {Qian}}\ and\ \bibinfo {author} {\bibfnamefont {T.~J.}\ \bibnamefont
  {Sejnowski}},\ }\href@noop {} {\bibfield  {journal} {\bibinfo  {journal}
  {Biol. Cyber.}\ }\textbf {\bibinfo {volume} {62}},\ \bibinfo {pages} {1}
  (\bibinfo {year} {1989})}\BibitemShut {NoStop}%
\bibitem [{\citenamefont {Mori}\ \emph {et~al.}(2007)\citenamefont {Mori},
  \citenamefont {Jerome},\ and\ \citenamefont {Peskin}}]{mori2007}%
  \BibitemOpen
  \bibfield  {author} {\bibinfo {author} {\bibfnamefont {Y.}~\bibnamefont
  {Mori}}, \bibinfo {author} {\bibfnamefont {J.~W.}\ \bibnamefont {Jerome}},\
  and\ \bibinfo {author} {\bibfnamefont {C.~S.}\ \bibnamefont {Peskin}},\
  }\href@noop {} {\bibfield  {journal} {\bibinfo  {journal} {Bull. Inst. Math.,
  Acad. Sin. (N.S.)}\ }\textbf {\bibinfo {volume} {2}},\ \bibinfo {pages} {367}
  (\bibinfo {year} {2007})}\BibitemShut {NoStop}%
\bibitem [{\citenamefont {Schuss}\ \emph {et~al.}(2001)\citenamefont {Schuss},
  \citenamefont {Nadler},\ and\ \citenamefont
  {Eisenberg}}]{schussEisenberg2001}%
  \BibitemOpen
  \bibfield  {author} {\bibinfo {author} {\bibfnamefont {Z.}~\bibnamefont
  {Schuss}}, \bibinfo {author} {\bibfnamefont {B.}~\bibnamefont {Nadler}},\
  and\ \bibinfo {author} {\bibfnamefont {R.~S.}\ \bibnamefont {Eisenberg}},\
  }\href@noop {} {\bibfield  {journal} {\bibinfo  {journal} {Phys. Rev. E}\
  }\textbf {\bibinfo {volume} {64}},\ \bibinfo {pages} {036116} (\bibinfo
  {year} {2001})}\BibitemShut {NoStop}%
\bibitem [{\citenamefont {singer}\ \emph {et~al.}(2008)\citenamefont {singer},
  \citenamefont {gillespie}, \citenamefont {norbury},\ and\ \citenamefont
  {eisenberg}}]{singer2008}%
  \BibitemOpen
  \bibfield  {author} {\bibinfo {author} {\bibfnamefont {A.}~\bibnamefont
  {singer}}, \bibinfo {author} {\bibfnamefont {D.}~\bibnamefont {gillespie}},
  \bibinfo {author} {\bibfnamefont {J.}~\bibnamefont {norbury}},\ and\ \bibinfo
  {author} {\bibfnamefont {R.~S.}\ \bibnamefont {eisenberg}},\ }\href@noop {}
  {\bibfield  {journal} {\bibinfo  {journal} {European Journal of Applied
  Mathematics}\ }\textbf {\bibinfo {volume} {19}},\ \bibinfo {pages}
  {541–560} (\bibinfo {year} {2008})}\BibitemShut {NoStop}%
\bibitem [{\citenamefont {Pods}\ \emph {et~al.}(2013)\citenamefont {Pods},
  \citenamefont {Schönke},\ and\ \citenamefont {Bastian}}]{pods2013}%
  \BibitemOpen
  \bibfield  {author} {\bibinfo {author} {\bibfnamefont {J.}~\bibnamefont
  {Pods}}, \bibinfo {author} {\bibfnamefont {J.}~\bibnamefont {Schönke}},\
  and\ \bibinfo {author} {\bibfnamefont {P.}~\bibnamefont {Bastian}},\
  }\href@noop {} {\bibfield  {journal} {\bibinfo  {journal} {Biophysical
  Journal}\ }\textbf {\bibinfo {volume} {105}},\ \bibinfo {pages} {242}
  (\bibinfo {year} {2013})}\BibitemShut {NoStop}%
\bibitem [{\citenamefont {Tuckwell}(1988)}]{tuckwell1988}%
  \BibitemOpen
  \bibfield  {author} {\bibinfo {author} {\bibfnamefont {H.~C.}\ \bibnamefont
  {Tuckwell}},\ }\href@noop {} {\emph {\bibinfo {title} {Introduction to
  theoretical neurobiology. Volume 1: Linear cable theory and dendritic
  structure}}}\ (\bibinfo  {publisher} {Cambridge University Press},\ \bibinfo
  {year} {1988})\BibitemShut {NoStop}%
\bibitem [{\citenamefont {Segev}\ and\ \citenamefont {Rall}(1988)}]{rallSegev}%
  \BibitemOpen
  \bibfield  {author} {\bibinfo {author} {\bibfnamefont {I.}~\bibnamefont
  {Segev}}\ and\ \bibinfo {author} {\bibfnamefont {W.}~\bibnamefont {Rall}},\
  }\href@noop {} {\bibfield  {journal} {\bibinfo  {journal} {J. Neurophysiol.}\
  }\textbf {\bibinfo {volume} {60}},\ \bibinfo {pages} {499} (\bibinfo {year}
  {1988})}\BibitemShut {NoStop}%
\bibitem [{\citenamefont {Savtchenko}\ \emph {et~al.}(2017)\citenamefont
  {Savtchenko}, \citenamefont {Poo},\ and\ \citenamefont
  {Rusakov}}]{rusakov2017}%
  \BibitemOpen
  \bibfield  {author} {\bibinfo {author} {\bibfnamefont {L.~P.}\ \bibnamefont
  {Savtchenko}}, \bibinfo {author} {\bibfnamefont {M.~M.}\ \bibnamefont
  {Poo}},\ and\ \bibinfo {author} {\bibfnamefont {D.~A.}\ \bibnamefont
  {Rusakov}},\ }\href@noop {} {\bibfield  {journal} {\bibinfo  {journal} {Nat.
  Rev. Neurosci.}\ }\textbf {\bibinfo {volume} {18}},\ \bibinfo {pages} {598}
  (\bibinfo {year} {2017})}\BibitemShut {NoStop}%
\bibitem [{\citenamefont {Cartailler}\ \emph {et~al.}(2018)\citenamefont
  {Cartailler}, \citenamefont {Kwon}, \citenamefont {Yuste},\ and\
  \citenamefont {Holcman}}]{cartailler2018neuron}%
  \BibitemOpen
  \bibfield  {author} {\bibinfo {author} {\bibfnamefont {J.}~\bibnamefont
  {Cartailler}}, \bibinfo {author} {\bibfnamefont {T.}~\bibnamefont {Kwon}},
  \bibinfo {author} {\bibfnamefont {R.}~\bibnamefont {Yuste}},\ and\ \bibinfo
  {author} {\bibfnamefont {D.}~\bibnamefont {Holcman}},\ }\href@noop {}
  {\bibfield  {journal} {\bibinfo  {journal} {Neuron}\ }\textbf {\bibinfo
  {volume} {97}},\ \bibinfo {pages} {1126} (\bibinfo {year}
  {2018})}\BibitemShut {NoStop}%
\bibitem [{\citenamefont {Safinya}\ and\ \citenamefont
  {Radler}(2021)}]{andelman}%
  \BibitemOpen
  \bibfield  {author} {\bibinfo {author} {\bibfnamefont {C.~R.}\ \bibnamefont
  {Safinya}}\ and\ \bibinfo {author} {\bibfnamefont {J.~E.}\ \bibnamefont
  {Radler}},\ }\href@noop {} {\emph {\bibinfo {title} {Handbook of Lipid
  Membranes: Molecular, Functional, and Materials Aspects (1st ed.)}}}\
  (\bibinfo  {publisher} {CRC Press},\ \bibinfo {year} {2021})\ p.\ \bibinfo
  {pages} {376}\BibitemShut {NoStop}%
\bibitem [{\citenamefont {Ben-Yaakov}\ \emph {et~al.}(2009)\citenamefont
  {Ben-Yaakov}, \citenamefont {Andelman}, \citenamefont {Harries},\ and\
  \citenamefont {Podgornik}}]{BenYaakov2009}%
  \BibitemOpen
  \bibfield  {author} {\bibinfo {author} {\bibfnamefont {D.}~\bibnamefont
  {Ben-Yaakov}}, \bibinfo {author} {\bibfnamefont {D.}~\bibnamefont
  {Andelman}}, \bibinfo {author} {\bibfnamefont {D.}~\bibnamefont {Harries}},\
  and\ \bibinfo {author} {\bibfnamefont {R.}~\bibnamefont {Podgornik}},\
  }\href@noop {} {\bibfield  {journal} {\bibinfo  {journal} {J. Phys.: Condens.
  Matter}\ }\textbf {\bibinfo {volume} {21}},\ \bibinfo {pages} {424106}
  (\bibinfo {year} {2009})}\BibitemShut {NoStop}%
\bibitem [{\citenamefont {Borukhov}\ \emph {et~al.}(2000)\citenamefont
  {Borukhov}, \citenamefont {Andelman},\ and\ \citenamefont
  {Orland}}]{orland2000}%
  \BibitemOpen
  \bibfield  {author} {\bibinfo {author} {\bibfnamefont {I.}~\bibnamefont
  {Borukhov}}, \bibinfo {author} {\bibfnamefont {D.}~\bibnamefont {Andelman}},\
  and\ \bibinfo {author} {\bibfnamefont {H.}~\bibnamefont {Orland}},\
  }\href@noop {} {\bibfield  {journal} {\bibinfo  {journal} {Electrochi. Acta}\
  }\textbf {\bibinfo {volume} {46}},\ \bibinfo {pages} {221} (\bibinfo {year}
  {2000})}\BibitemShut {NoStop}%
\bibitem [{\citenamefont {Garcia}\ \emph {et~al.}(2019)\citenamefont {Garcia},
  \citenamefont {Bartol}, \citenamefont {Phan}, \citenamefont {Bushong},
  \citenamefont {Perkins}, \citenamefont {Sejnowski}, \citenamefont
  {Ellisman},\ and\ \citenamefont {Skupin}}]{garcia}%
  \BibitemOpen
  \bibfield  {author} {\bibinfo {author} {\bibfnamefont {G.~C.}\ \bibnamefont
  {Garcia}}, \bibinfo {author} {\bibfnamefont {T.~M.}\ \bibnamefont {Bartol}},
  \bibinfo {author} {\bibfnamefont {S.}~\bibnamefont {Phan}}, \bibinfo {author}
  {\bibfnamefont {E.~A.}\ \bibnamefont {Bushong}}, \bibinfo {author}
  {\bibfnamefont {G.}~\bibnamefont {Perkins}}, \bibinfo {author} {\bibfnamefont
  {T.~J.}\ \bibnamefont {Sejnowski}}, \bibinfo {author} {\bibfnamefont {M.~H.}\
  \bibnamefont {Ellisman}},\ and\ \bibinfo {author} {\bibfnamefont
  {A.}~\bibnamefont {Skupin}},\ }\href@noop {} {\bibfield  {journal} {\bibinfo
  {journal} {Sci. Rep.}\ }\textbf {\bibinfo {volume} {9}},\ \bibinfo {pages}
  {18306} (\bibinfo {year} {2019})}\BibitemShut {NoStop}%
\bibitem [{\citenamefont {Bazant}\ \emph {et~al.}(2004)\citenamefont {Bazant},
  \citenamefont {Thornton},\ and\ \citenamefont {Ajdari}}]{bazant}%
  \BibitemOpen
  \bibfield  {author} {\bibinfo {author} {\bibfnamefont {M.~Z.}\ \bibnamefont
  {Bazant}}, \bibinfo {author} {\bibfnamefont {K.}~\bibnamefont {Thornton}},\
  and\ \bibinfo {author} {\bibfnamefont {A.}~\bibnamefont {Ajdari}},\
  }\href@noop {} {\bibfield  {journal} {\bibinfo  {journal} {Phys. Rev. E}\
  }\textbf {\bibinfo {volume} {70}},\ \bibinfo {pages} {021506} (\bibinfo
  {year} {2004})}\BibitemShut {NoStop}%
\bibitem [{com(n 61)}]{comsol}%
  \BibitemOpen
  \href@noop {} {\emph {\bibinfo {title} {COMSOL Multiphysics}}} (\bibinfo
  {year} {Version 6.1}),\ \bibinfo {note}
  {\url{https://www.comsol.com/release/6.1}}\BibitemShut {NoStop}%
\bibitem [{\citenamefont {Cartailler}\ \emph {et~al.}(2017)\citenamefont
  {Cartailler}, \citenamefont {Schuss},\ and\ \citenamefont
  {Holcman}}]{cartailler2017jns}%
  \BibitemOpen
  \bibfield  {author} {\bibinfo {author} {\bibfnamefont {J.}~\bibnamefont
  {Cartailler}}, \bibinfo {author} {\bibfnamefont {Z.}~\bibnamefont {Schuss}},\
  and\ \bibinfo {author} {\bibfnamefont {D.}~\bibnamefont {Holcman}},\
  }\href@noop {} {\bibfield  {journal} {\bibinfo  {journal} {J. Nonlinear
  Sci.}\ }\textbf {\bibinfo {volume} {27}},\ \bibinfo {pages} {1971} (\bibinfo
  {year} {2017})}\BibitemShut {NoStop}%
\bibitem [{\citenamefont {Struik}(1988)}]{struik1988}%
  \BibitemOpen
  \bibfield  {author} {\bibinfo {author} {\bibfnamefont {D.~J.}\ \bibnamefont
  {Struik}},\ }\href@noop {} {\emph {\bibinfo {title} {Lectures on classical
  differential geometry (2nd ed.)}}},\ Dover Books Adv. Math.\ (\bibinfo
  {publisher} {New York: Dover Publications, Inc.},\ \bibinfo {year}
  {1988})\BibitemShut {NoStop}%
\bibitem [{\citenamefont {Boahen}\ and\ \citenamefont
  {Doyon}(2020)}]{doyon2020}%
  \BibitemOpen
  \bibfield  {author} {\bibinfo {author} {\bibfnamefont {F.}~\bibnamefont
  {Boahen}}\ and\ \bibinfo {author} {\bibfnamefont {N.}~\bibnamefont {Doyon}},\
  }\href@noop {} {\bibfield  {journal} {\bibinfo  {journal} {J. Math. Biol.}\
  }\textbf {\bibinfo {volume} {81}},\ \bibinfo {pages} {517} (\bibinfo {year}
  {2020})}\BibitemShut {NoStop}%
\bibitem [{\citenamefont {Biess}\ \emph {et~al.}(2007)\citenamefont {Biess},
  \citenamefont {Korkotian},\ and\ \citenamefont
  {Holcman}}]{biess2007diffusion}%
  \BibitemOpen
  \bibfield  {author} {\bibinfo {author} {\bibfnamefont {A.}~\bibnamefont
  {Biess}}, \bibinfo {author} {\bibfnamefont {E.}~\bibnamefont {Korkotian}},\
  and\ \bibinfo {author} {\bibfnamefont {D.}~\bibnamefont {Holcman}},\
  }\href@noop {} {\bibfield  {journal} {\bibinfo  {journal} {Phys. Rev. E}\
  }\textbf {\bibinfo {volume} {76}},\ \bibinfo {pages} {021922} (\bibinfo
  {year} {2007})}\BibitemShut {NoStop}%
\bibitem [{\citenamefont {Cheviakov}\ \emph {et~al.}(2010)\citenamefont
  {Cheviakov}, \citenamefont {Ward},\ and\ \citenamefont
  {Straube}}]{cheviakov2010}%
  \BibitemOpen
  \bibfield  {author} {\bibinfo {author} {\bibfnamefont {A.~F.}\ \bibnamefont
  {Cheviakov}}, \bibinfo {author} {\bibfnamefont {M.~J.}\ \bibnamefont
  {Ward}},\ and\ \bibinfo {author} {\bibfnamefont {R.}~\bibnamefont
  {Straube}},\ }\href@noop {} {\bibfield  {journal} {\bibinfo  {journal}
  {Multiscale Model. Simul.}\ }\textbf {\bibinfo {volume} {8}},\ \bibinfo
  {pages} {836} (\bibinfo {year} {2010})}\BibitemShut {NoStop}%
\end{thebibliography}
%
\end{document}